\documentclass[12pt]{article}
\usepackage{latexsym}
\usepackage{amsfonts,amssymb,amsmath} 
\usepackage{mathtools}
\usepackage[a4paper,vmargin=3cm,hmargin=2.1cm]{geometry}
\usepackage{tikz}
\usepackage{hyperref}
\usepackage{lmodern}

\newcommand{\double}[1]{\mathbb{#1}}

\newcommand{\rr}{\double{R}}

\newcommand{\diff}{\text{\textsl{Diff}}}
\newcommand{\iso}{\text{\textsl{Iso}}}
\DeclareMathOperator{\Ar}{Ar}

\newcommand{\de}{\hbox{\rm{d}}}

\newcommand{\pa}{\partial}

\newcommand{\lb}{\left[}
\newcommand{\rb}{\right]}
\newcommand{\lp}{\left(}
\newcommand{\rp}{\right)}

\newcommand{\dpp}{\vcentcolon}
\newcommand{\bb}{\begin{eqnarray}}
\newcommand{\ee}{\end{eqnarray}}
\newcommand{\eee}{\nonumber\end{eqnarray}}
\newcommand{\qq}{\quad}

\begin{document}

\thispagestyle{empty}

\begin{center}
${}$
\vspace{3cm}

{\Large\textbf{Maximal symmetry at the speed of light}} \\

\vspace{2cm}

{\large
Andr\'e Tilquin\footnote{Tsinghua Center for Astrophysics, Tsinghua University, Beijing, China \\\indent\qq 
CPPM, Aix-Marseille University, 13288 Marseille, France\\\indent\qq tilquin@cppm.in2p3.fr }, Thomas Sch\"ucker\footnote{
Aix-Marseille University, CNRS UMR 7332, CPT, 13288 Marseille, France\\\indent\qq
Universit\'e de Toulon, CNRS UMR 7332, CPT, 83957 La Garde, France
\\\indent\qq
thomas.schucker@gmail.com } }

\vspace{3cm}

{\large\textbf{Abstract}}
\end{center}
We propose a relativistic version of the cosmological principle and confront it to the Hubble diagram of
supernovae and other probes.

\vspace{5.6cm}

\vskip 1truecm

\noindent CPT-2011/P004\\
\noindent 1104.0160

\section{Introduction}

Differential equations governing the time evolution of a field (of competitors) is one of the main paradigms in physics. To obtain a (locally) unique solution we must specify initial data of the field on a Cauchy surface (the starting line). Even for relativistic wave equations, we are used to taking the Cauchy surface space-like. In the absence of super-luminal signals, we would prefer to remain humble and use light-like `Cauchy' surfaces. 
In other words we acknowledge several obstacles to the idealisation behind  space-like Cauchy surfaces: {\it (i)} The starting line is too long and in the short period allocated to us since de-ionisation, we cannot observe it entirely using photons, that travel only at the speed of light. {\it (ii)} The competitors do not remain at rest on their marks prior to kick-off. You know how complicated a sailing regatta is at the starting line. Now add to it another complication: The jury is only able to locate the boats waiting around the starting line with go-betweens traveling at speeds comparable to the speed of the waiting boats.

We do not see how to make sense out of this complicated space-like initial value problem in cosmology without superluminal go-betweens. On the other hand, the Cauchy problem of relativistic equations
 with light-like surfaces is much more involved than with space-like ones \cite{fried}, but this is not the subject of the present paper concerned with the Killing equation on certain light-like surfaces.
 
 To set the stage we will give a pedestrian derivation of the Robertson-Walker metrics of negative, vanishing and positive curvatures. It is naive, but can be generalised from space-like surfaces of `simultaneity' to the light-like ones of our past light-cones. The ensuing metrics together with Einstein's equations are then confronted to the Hubble diagram of supernovae and other probes.
  
 \section{ Isometry groups}
 
 Let us recall briefly the basic definitions of isometries and a few important theorems.
 
 Consider a (pseudo) Riemannian manifold $(M,g)$ with arbitrary signature and of dimension $d$. In coordinates $x^\mu $ its metric $g$ is expressed by its metric tensor, $\de \tau^2=g_{\mu \nu}(x)\,\de x^\mu \,\de x^\nu.$ Let $f $ be a diffeomorphism on $M$. They form a group written $\diff(M)$ which is infinite dimensional and not a Lie group. We denote the Jacobian of $f $ in coordinates by
 \bb {\Lambda ^{\bar \mu }}_\mu(x) \dpp= 
\,\frac{\pa f ^{\bar \mu }\ \ }{\ \ \pa x ^\mu }\, (x).
\ee
By definition $f $ is an isometry  if
\bb
g_{\mu \nu}(x) ={\left( \Lambda ^T\right)_\mu }^{\bar\mu}(x)\, g_{\bar\mu \bar\nu} (f (x))\,{\Lambda ^{\bar \nu }}_\nu(x)\,\label{iso}
\ee
in all charts.
The isometries form a subgroup $\iso(M,g)\subset 
\diff(M)$. The main theorem says that the isometry group is a Lie group of finite dimension less than or equal to $d(d+1)/2$. The (pseudo) Riemannian manifold is said to be maximally symmetric if its isometry group is of maximal dimension $d(d+1)/2$. In dimension 1+3 there are three types of maximally symmetric spaces,  anti de Sitter, Minkowski and de Sitter spaces. They solve the vacuum Einstein equations with negative, vanishing and positive cosmological constant, $\Lambda = 3\sigma k^2$ in the notations defined below. Similarily in dimensions 0+3, we have the maximally symmetric spaces: the pseudo-spheres, Euclidean $\rr^3$ and the spheres.

Equation (\ref{iso}) is a functional differential equation and it is practical to turn it into a differential equation by  considering infinitesimal diffeomorphisms.
Upon linearisation $f (x)=x+\xi (x)+o(\xi ^2)$, where $\xi=\xi ^\alpha \,\pa/\pa x^\alpha  $ is a vector field, equation (\ref{iso}) becomes the Killing equation:
\bb 
\xi ^\alpha \,\frac{\pa}{\pa x^\alpha }\, g_{\mu \nu}+\,\frac{\pa \xi ^{\bar\mu }}{\pa x^\mu }\, g_{\bar\mu \nu}+\,\frac{\pa \xi ^{\bar\nu }}{\pa x^\nu }\, g_{\mu \bar\nu}=0.\label{kill}\ee
It serves two purposes: if the generators of the isometry group are given, the Killing equation is a first order differential equation in the metric tensor. Its solutions tell us what metrics are allowed by the given symmetries. On the other hand for a given metric, the Killing equation is a first order differential equation in the vector fields. Its solutions generate the isometry group. In both cases there are global issues concerning the patching up of charts that we will happily ignore in this paper.
 
 \section{ A pedestrian derivation of the Robertson-Walker metrics}
 
 In this section we recall a local derivation of the Robertson-Walker metrics using maximal symmetry on spacetime and on surfaces of `simultaneity'. For a thermodynamic derivation the interested reader is referred to Jean-Marie Souriau \cite{sour}.
 
 Our starting points are the maximally symmetric spacetimes: anti de Sitter $\sigma =-1$, Minkowski $\sigma =0$ and de Sitter $\sigma =+1$ in polar coordinates $(t,\,r,\, \theta ,\,\varphi )$, where $t$ is cosmic time and $r$ is the geodesic `distance'. (It is usually denoted by $\chi $, but we will use the letter $\chi $ later for a different coordinate.) 
 We will use the following auxiliary functions:
 \bb
 s(r)\dpp =
 \left\{ \begin{array}{ll}
 \sinh(kr)/k & \sigma =-1\\
 r&\sigma =0\\
 \sin(kr)/k&\sigma =+1\qq ,
 \end{array}\right. 
 \ee
 where $k$ is positive and has the dimension of an inverse meter, and
 \bb
 c(r)\dpp =
 \left\{ \begin{array}{ll}
 \cosh(kr) & \sigma =-1\\
 1&\sigma =0\\
 \cos(kr)&\sigma =+1\qq .
 \end{array}\right. 
 \ee
  Note the continuity properties
 \bb 
 \lim_{k\to 0}s_{\sigma =-1}=s_{\sigma =0}, &&  \lim_{k\to 0}s_{\sigma =+1}=s_{\sigma =0},\\[1mm]
 \lim_{k\to 0}c_{\sigma =-1}=c_{\sigma =0}, &&  \lim_{k\to 0}c_{\sigma =+1}=c_{\sigma =0}.
 \ee
 The following relations will be useful:
 \bb
 \sigma k^2s^2+c^2=1,\qq s'=c,\qq c'=-\sigma k^2s,\qq (c/s)'=-1/s^2,\qq (s/c)'=1/c^2.
 \ee
 Likewise we define:
\bb
 \bar s(t)\dpp =
 \left\{ \begin{array}{ll}
 \sin(kt)/k & \sigma =-1\\
 t&\sigma =0\\
 \sinh(kt)/k&\sigma =+1
 \end{array}\right. ,
 &&
 \bar c(t)\dpp =
 \left\{ \begin{array}{ll}
 \cos(kt) & \sigma =-1\\
 1&\sigma =0\\
 \cosh(kt)&\sigma =+1
 \end{array}\right. ,
 \ee
with similar relations. Now we can write down the metrics of our maximally symmetric spacetimes:
\bb
\de \tau^2=\de t^2-\bar c^2(t)\,\lb
\de r^2 + s^2(r)\,\de\theta ^2+s^2(r)\sin^2\theta \,\de \varphi ^2 \rb \label{max}
\ee
with the ten Killing vectors, three rotations, three space `translations', three Lorentz boosts, and a time `translation':
\bb &\begin{array}{ll}
R_x= -\sin\varphi \,\frac{\pa}{\pa\theta }\, -\,\frac{\cos\theta }{\sin\theta }\, \cos\varphi \,\frac{\pa}{\pa\varphi  }\,,&
T_x=\sin\theta \,\cos\varphi \,\frac{\pa}{\pa r}\,+
\,\frac{c}{s}\, \cos\theta\, \cos\varphi \,\frac{\pa}{\pa\theta }\,-\,\frac{c}{s}\,\frac{\sin\varphi }{\sin\theta }\, \frac{\pa}{\pa\varphi  }\,,\\[3mm]
R_y= +\cos\varphi \,\frac{\pa}{\pa\theta }\, -\,\frac{\cos\theta }{\sin\theta }\, \sin\varphi \,\frac{\pa}{\pa\varphi  }\,,&
T_y=\sin\theta \,\sin\varphi \,\frac{\pa}{\pa r}\,+
\,\frac{c}{s}\, \cos\theta\, \sin\varphi \,\frac{\pa}{\pa\theta }\,+\,\frac{c}{s}\,\frac{\cos\varphi }{\sin\theta }\, \frac{\pa}{\pa\varphi  }\,,\\[3mm]
R_z= \qq\qq\qq\qq\qq\qq\qq\qq\qq\qq \,\frac{\pa}{\pa\varphi  }\,,&
T_z=\cos\theta\, \ \ \cdot\,\ \  \frac{\pa}{\pa r}\,-
\,\frac{c}{s}\, \sin\theta\, \ \ \cdot\,\ \  \frac{\pa}{\pa\theta }\,,
\end{array}&\nonumber\\[3mm]
&\begin{array}{l}
L_x= s\,\sin\theta \,\cos\varphi \,\frac{\pa}{\pa t}\,+
\,\frac{\bar s}{\bar c}\,c\, \sin\theta\, \cos\varphi \,\frac{\pa}{\pa r }\,
+
\,\frac{\bar s}{\bar c}\,\frac{1}{s}\,  \cos\theta\, \cos\varphi \,\frac{\pa}{\pa \theta }\,
-\,\frac{\bar s}{\bar c}\,\frac{1}{s}\,\frac{\sin\varphi }{\sin\theta }\, \frac{\pa}{\pa\varphi  }\,,\\[3mm]
L_y= s\,\sin\theta \,\sin\varphi \,\frac{\pa}{\pa t}\,+
\,\frac{\bar s}{\bar c}\,c\, \sin\theta\, \sin\varphi \,\frac{\pa}{\pa r }\,
+
\,\frac{\bar s}{\bar c}\,\frac{1}{s}\,  \cos\theta\, \sin\varphi \,\frac{\pa}{\pa \theta }\,
+\,\frac{\bar s}{\bar c}\,\frac{1}{s}\,\frac{\cos\varphi }{\sin\theta }\, \frac{\pa}{\pa\varphi  }\,,\\[3mm]
L_z= s\,\cos\theta \, \ \ \cdot\,\ \  \frac{\pa}{\pa t}\,+
\,\frac{\bar s}{\bar c}\,c\, \cos\theta\, \ \ \cdot\,\ \  \frac{\pa}{\pa r }\,
-
\,\frac{\bar s}{\bar c}\,\frac{1}{s}\,  \sin\theta\, \ \ \cdot\,\ \  \frac{\pa}{\pa \theta }\,,
\end{array}&
\nonumber\\[3mm]
&\begin{array}{l}
T_t= c\,\frac{\pa}{\pa t}\,-
\sigma k^2
\,\frac{\bar s}{\bar c}\,s \,\frac{\pa}{\pa r }\,.
\end{array}&
\ee
As a check, let us compute a few Lie brackets:
\bb \begin{array}{llll}
[R_x,R_y]=-R_z,& [R_x,T_z]=T_y,& [T_y,T_z]=-\sigma k^2 R_x,
& [R_z,T_z]=0,\\
{[L_y,L_z]=R_x,}& [R_x,L_z]=L_y,&
[R_z,L_z]=0,& [T_x,L_z]=0,\\
 {[T_z,L_z]=T_t,}
& [R_z,T_t]=0,& [T_z,T_t]=-\sigma k^2L_z,& [L_z,T_t]=-T_z.
\end{array}
\label{check}
\ee
These Killing vectors generate the identity components of the isometry groups, the anti de Sitter group $O(2,3)$ for $\sigma =-1$, the Poincar\'e group $O(1,3)\ltimes \rr^4$ for $\sigma =0$ and the de Sitter group $O(1,4)$ for $\sigma =1$.

Maximally symmetric spacetimes are too rigid to match our observations of the universe. The conventional way to relax the symmetry is to only postulate maximal symmetry on space-like 3-surfaces of `simultaneity', $t=t_0$. These are only invariant under the three rotations $R_x,\,R_y,\,R_z$  and the three `translations' $T_x,\,T_y,\,T_z$ generating the identity components of the Lorentz group $O(1,3)$ for  $\sigma =-1$, of the 3-dimensional Euclidean group $O(3)\ltimes\rr^3$ for  $\sigma =0$ and of $O(4)$ for  $\sigma =1$. They are the isometry groups of pseudo-spheres, the Euclidean 3-space and spheres of radius $\bar c(t_0)/k$. Note a dangerous trap with pseudo-spheres. Their isometry groups are isomorphic to the Lorentz group. Its elements are genuine rotations, but the Lorentz transformations are fake, for there is no time. They are fake translations, fake because they do not commute. 

 The most general spacetime metrics solving the Killing equations (\ref{kill}) for the six Killing vectors $R_x,\,R_y,\,R_z,\ T_x,\,T_y,\,T_z$ are the Robertson-Walker metrics
\bb \de\tau^2=b^2(t)\,\de t^2-a^2(t)\lb
\de r^2 + s^2(r)\,\de\theta ^2+s^2(r)\sin^2\theta \,\de \varphi ^2 \rb, \label{rw}
\ee
with positive functions $a(t)$ and $b(t)$. By a transformation of the time coordinate we may achieve $b\equiv 1$, in  `cosmic time', that we still denote by $t$, or we may achieve $b\equiv a$, in `conformal time', that we denote by $\eta$. The link between cosmic  time and conformal time is given by $\de \eta/\de t=1/a(t)$. The Robertson-Walker spacetimes are foliated by the time coordinate with 3-dimensional, space-like leaves of maximal symmetry, pseudo-spheres, Euclidean 3-space and spheres.

Note that in all three cases the time axis is automatically `perpendicular' to the leaves without invoking a `Weyl principle'. This comes from the embedding of the six 3-dimensional Killing vectors (on the leaves) into spacetime, an embedding that these vectors inherit from the maximally symmetric spacetimes we started from. For a recent analysis of `Weyl's principle' and an example relaxing it, see Marinoni \& Steigerwald \cite{ms}.

\section{ Light-like leaves}

Our task is to exhibit a spacetime foliated by 3-dimensional, light-like leaves, our past light-cones, with `maximal symmetry'. We use the quotation marks because the main theorem above fails for general degenerate metrics: They may well have infinite dimensional isometry groups. Take the direct product of the real line with vanishing `metric' and the round 2-sphere. Its isometry group is $\diff(\rr)\times O(3)$. We find it encouraging that this problem will not appear when we repeat the above steps for the three families of maximally symmetric spacetimes (\ref{max}), replacing however the space-like leaves  of `simultaneity' by the light-like leaves of our past light-cones. 

Of course we will switch to light-cone coordinates in conformal time:
\bb
\begin{array}{lll}
 u \dpp = {\textstyle\frac{1}{\sqrt 2}}(\eta + r),&
 \eta={\textstyle\frac{1}{\sqrt 2}}(u +v),&
 \frac{\pa}{\pa t}=\frac{1}{\sqrt 2}\frac{1}{a}\lp\frac{\pa}{\pa u}+\frac{\pa}{\pa v}\rp,\\[3mm]
 v \dpp = {\textstyle\frac{1}{\sqrt 2}}(\eta - r),&
 r={\textstyle\frac{1}{\sqrt 2}}(u-v),&
 \frac{\pa}{\pa r}=\frac{1}{\sqrt 2}\lp\frac{\pa}{\pa u}-\frac{\pa}{\pa v}\rp.
 \end{array}
 \ee
In these coordinates the maximally symmetric metrics (\ref{max}), $a(t)=\bar c(t)$, read:
\bb
\de \tau^2=\bar c^2(t({\textstyle\frac{1}{\sqrt 2}}(u +v)))\lb 2\,\de u\,\de v
- s^2({\textstyle\frac{1}{\sqrt 2}}(u-v))\,\de\theta ^2-s^2({\textstyle\frac{1}{\sqrt 2}}(u-v))\sin^2\theta \,\de \varphi ^2 \rb \label{max2},
\ee
and their Lorentz transformations are:
\bb
\begin{array}{l}
L_x= 
{\textstyle\frac{1}{\sqrt 2}}\,\,\frac{1}{\bar c}\, (s+\bar sc)
\,\sin\theta \,\cos\varphi \,\frac{\pa}{\pa u}\,+
{\textstyle\frac{1}{\sqrt 2}}\,\,\frac{1}{\bar c}\, (s-\bar sc)
\,\sin\theta \,\cos\varphi \,\frac{\pa}{\pa v}\\[2mm]
\hspace{9cm} +
\,\frac{\bar s}{\bar c}\,\frac{1}{s}\,  \cos\theta\, \cos\varphi \,\frac{\pa}{\pa \theta }\,
-\,\frac{\bar s}{\bar c}\,\frac{1}{s}\,\frac{\sin\varphi }{\sin\theta }\, \frac{\pa}{\pa\varphi  }\,,\\[3mm]
L_y= 
{\textstyle\frac{1}{\sqrt 2}}\,\,\frac{1}{\bar c}\, (s+\bar sc)
\,\sin\theta \,\sin\varphi \,\frac{\pa}{\pa u}\,+
{\textstyle\frac{1}{\sqrt 2}}\,\,\frac{1}{\bar c}\, (s-\bar sc)
\,\sin\theta \,\sin\varphi \,\frac{\pa}{\pa v}\\[2mm]
\hspace{9cm} 
+
\,\frac{\bar s}{\bar c}\,\frac{1}{s}\,  \cos\theta\, \sin\varphi \,\frac{\pa}{\pa \theta }\,
+\,\frac{\bar s}{\bar c}\,\frac{1}{s}\,\frac{\cos\varphi }{\sin\theta }\, \frac{\pa}{\pa\varphi  }\,,\\[3mm]
L_z= {\textstyle\frac{1}{\sqrt 2}}\,\,\frac{1}{\bar c}\, (s+\bar sc)\,\cos\theta \, \ \ \cdot\,\ \  \frac{\pa}{\pa u}\,+
{\textstyle\frac{1}{\sqrt 2}}\,\,\frac{1}{\bar c}\, (s-\bar sc)
\, \cos\theta\, \ \ \cdot\,\ \  \frac{\pa}{\pa v }\\[2mm]
\hspace{9cm} 
-
\,\frac{\bar s}{\bar c}\,\frac{1}{s}\,  \sin\theta\, \ \ \cdot\,\ \  \frac{\pa}{\pa \theta }\,,
\end{array}\label{boosts1}
\ee
where it is understood that the arguments $t$ and $r$ are replaced by $t=t(\eta),\ \eta={\textstyle\frac{1}{\sqrt 2}}(u +v),
\ r={\textstyle\frac{1}{\sqrt 2}}(u-v).$
On the past light-cones $u=u_0$, the 4-metrics reduce to the {\it degenerate} 3-metrics:
\bb
\de \nu^2= -\tilde a^2(v)\lb 
 \de\theta ^2+\sin^2\theta \,\de \varphi ^2 \rb \label{deg3},\qq \tilde a(v)\dpp =\bar c(t({\textstyle\frac{1}{\sqrt 2}}(u_0 +v)))\,s({\textstyle\frac{1}{\sqrt 2}}(u_0-v)). \label{a}
\ee
 To compute its Killing vectors we need the derivative of $\tilde a$:
\bb
\tilde a'=\,\frac{\de\tilde a(v)}{\de v}\, ={\textstyle\frac{1}{\sqrt 2}} \,\bar c(t)\lb\sigma k^2\bar s(t)\,s(r)-c(r)\rb,\qq
t=t(\eta),\ \eta={\textstyle\frac{1}{\sqrt 2}}(u_0 +v),
\ r={\textstyle\frac{1}{\sqrt 2}}(u_0-v),\ee
and we find that the Killing vectors form a 6-dimensional Lie algebra spanned by the three rotations $R_x,\,R_y,\,R_z$  and the three Lorentz transformations $\tilde L_x,\,\tilde L_y,\,\tilde L_z$:
\bb
\begin{array}{l}
\tilde L_x= 
-\,\frac{\tilde a}{\tilde a'} 
\,\sin\theta \,\cos\varphi \,\frac{\pa}{\pa v}\,
-\,
 \cos\theta\, \cos\varphi \,\frac{\pa}{\pa \theta }\,
+\,\frac{\sin\varphi }{\sin\theta }\, \frac{\pa}{\pa\varphi  }\,,\\[3mm]
\tilde L_y= 
-\,\frac{\tilde a}{\tilde a'} 
\,\sin\theta \,\sin\varphi \,\frac{\pa}{\pa v}\,
-\,
 \cos\theta\, \sin\varphi \,\frac{\pa}{\pa \theta }\,
-\,\frac{\cos\varphi }{\sin\theta }\, \frac{\pa}{\pa\varphi  }\,,\\[3mm]
\tilde L_z= 
-\,\frac{\tilde a}{\tilde a'} 
\,\cos\theta \,\ \ \cdot\ \ \,\frac{\pa}{\pa v}\,
+\,
 \sin\theta\,\ \ \cdot\ \ \, \frac{\pa}{\pa \theta }
\,.\end{array}\label{boosts2}\ee
 Indeed, we get the same commutation relations as in (\ref{check}), ${[\tilde L_y,\tilde L_z]=R_x,}\  [R_x,\tilde L_z]=\tilde L_y,$ $ 
[R_z,\tilde L_z] =0$, and the Killing vectors generate the identity component of the Lorentz group $O(1,3)$. These commutation relations continue to hold for arbitrary functions $\tilde a (v)$ with nowhere vanishing derivative, not necessarily of the form in equation (\ref{a}). It is worth noting that while in the case of surfaces of `simultaneity' the isometry groups are different for the three families of metrics, $\sigma =-1,\,0,\,+1$, these three groups are isomorphic in the case of light-cones. This is remarkable because we are talking about light-cones in spacetime, not in tangent space, and they are cones only for $\sigma =0$. For $\sigma =1$ the past light-cones even have two singularities. Note also that this time the Lorentz transformations are not fake, they are genuine boosts. In the case $\sigma =0$, the Lorentz boosts with tildes in equations (\ref{boosts2}) are the restrictions of the Lorentz boosts in equations (\ref{boosts1})  to the light-cone $u=u_0$ after the shift $u\rightarrow u-u_0$ and $v\rightarrow v-u_0$.

Now that we know the isometry group on the leaves, we compute -- as in the last section -- the most general metric on the foliated spacetimes compatible with the Killing vectors, $R_x,\,R_y,\,R_z,\ L_x,\,L_y,\,L_z$.  

To this end it is convenient to work in still another coordinate system:
\bb \chi \dpp =\sqrt{\bar s^2(t)-\bar c^2(t)\,s^2(r)}
=\sqrt{\bar s^2(t)\,c^2(r)-s^2(r)},\qq
\psi \dpp =\Ar\tanh\,\frac{\bar c(t)\,s(r)}{\bar s(t)}\, ,\label{cotr}\ee
with Jacobian
\bb J\dpp =
\begin{pmatrix}
{\pa\chi }/{\pa t} &{\pa\psi  }/{\pa t}\\[2mm]
{\pa\chi }/{\pa r} &{\pa\psi  }/{\pa r}
\end{pmatrix} 
=
\begin{pmatrix}
{\bar s\bar cc^2 }/{\chi } &-s/\chi ^2\\[2mm]
-\bar c^2sc/\chi  &\bar s\bar cc/\chi ^2
\end{pmatrix} \qq \text{and}\qq
J^{-1}=
\begin{pmatrix}
{\bar s/\bar c }/{\chi } &-s/(\bar c^2c\chi) \\[2mm]
s  &\bar sc/\bar c
\end{pmatrix}.
\ee
Then with
\bb
\,\frac{\pa}{\pa \psi }\,=\,\frac{\pa t}{\pa \psi }\,\frac{\pa}{\pa t }\,+\,\frac{\pa r}{\pa \psi }\,\frac{\pa}{\pa r }\,=
s\,\frac{\pa}{\pa t }\,+\,\frac{\bar s c}{\bar c }\,\frac{\pa}{\pa r }
\ee
we get
\bb
L_z= \cos\theta \, \frac{\pa}{\pa \psi }\,-
\,\coth \psi \,  \sin\theta\, \frac{\pa}{\pa \theta }\,.\ee
We find it amazing that in these coordinates the boosts are independent of $\sigma $.

It is well-known that the most general solution of the Killing equation with respect to the rotations $R_x,\,R_y,\,R_z$ is
\bb \de\tau^2=B\,\de \chi ^2
+2S\,\de\chi \,\de\psi -A\,
\de \psi ^2 - C\,\de\theta ^2-C\sin^2\theta \,\de \varphi ^2 , 
\ee 
where $A,\,B,\,C$ and $S$ are functions of $\chi $ and $\psi $. Since all boosts can be obtained by commuting rotations with $L_z$, the functions $A,\,B,\,C$ and $S$ are determined from the Killing equation for $L_z$ alone. This calculation is straight-forward and  yields the Robertson-Walker metrics with $\sigma =-1$: $S=0$, $C=a^2(\chi )\,\sinh^2\psi $, $B=b^2(\chi )$ and $A=a^2(\chi )$.
 
 The coordinate transformation (\ref{cotr}) has a singularity at $\chi =0$. To be sure that this singularity does not spoil our conclusion, we rewrite the obtained metric,
 \bb \de\tau^2=b^2(\chi )\,\de \chi ^2
 -a^2(\chi )\,
\de \psi ^2 - a^2(\chi )\,\sinh^2\psi \,\de\theta ^2-a^2(\chi )\,\sinh^2\psi \sin^2\theta \,\de \varphi ^2 , 
\ee 
in the light-cone coordinates $u$ and $v$. To simplify we choose the embedding into Minkowski space and we choose $\chi $ such that $a^2=2uv\,b^2$. Then we get
\bb
\de \tau^2=b^2(\sqrt{2uv})\lb 2\,\de u\,\de v
- {\textstyle\frac{1}{ 2}}(u-v)^2)\,\de\theta ^2-{\textstyle\frac{1}{ 2}}(u-v)^2\sin^2\theta \,\de \varphi ^2 \rb .
\ee 
This metric has light-like leaves $u=u_0$ with maximal symmetry. The isometry groups are the Lorentz group (at least its connected component) as long as $\de/\de v\,[b(\sqrt{2u_0v})\,(u_0-v)]\not=0$.

Our little calculation indicates that the pseudo-spheres are universal with respect to the relativistic version of the cosmological principle. But we cannot exclude that are other metrics admitting a foliation with maximally symmetric, light-like leaves; metrics that cannot be obtained from maximally symmetric spacetimes. Any such metric, of course, would be extremely interesting to be tested against cosmological observations.

Let us state this universality of pseudo-spheres differently.
The Robertson-Walker metrics with $\sigma =-1$ admit two foliations: The first, $\psi =\psi _0$, has space-like leaves with maximal symmetry and the fake Lorentz group. The second, $u=u_0$, has light-like leaves with maximal symmetry and the genuine Lorentz group. We
conjecture that the other two families, with spheres and planes, do not admit the second type of foliation. Before mathematicians prove or disproves this conjecture, let us see what cosmological observations tell us. Of course we start with the cleanest (because coordinate independent) test, the Hubble diagram.

\section{Data analysis}

We have seen above that the metrics associated with maximal symmetry on
past light-cones are the Robertson-Walker metrics with $\sigma=-1$. 
Consequently, the Friedmann equations can be used with the restriction
 of a positive curvature density parameter $\Omega_k$ and we use the published results on supernovae directly to test the
value of $\Omega_k$.

The most up-to-date results for supernovae are coming from the SCP
compilation experiment \cite{suzuki} and the latest SNLS 3 years data \cite{snls}.

\begin{figure}[h]
\begin{center}
\includegraphics[width=7cm, height=6cm]{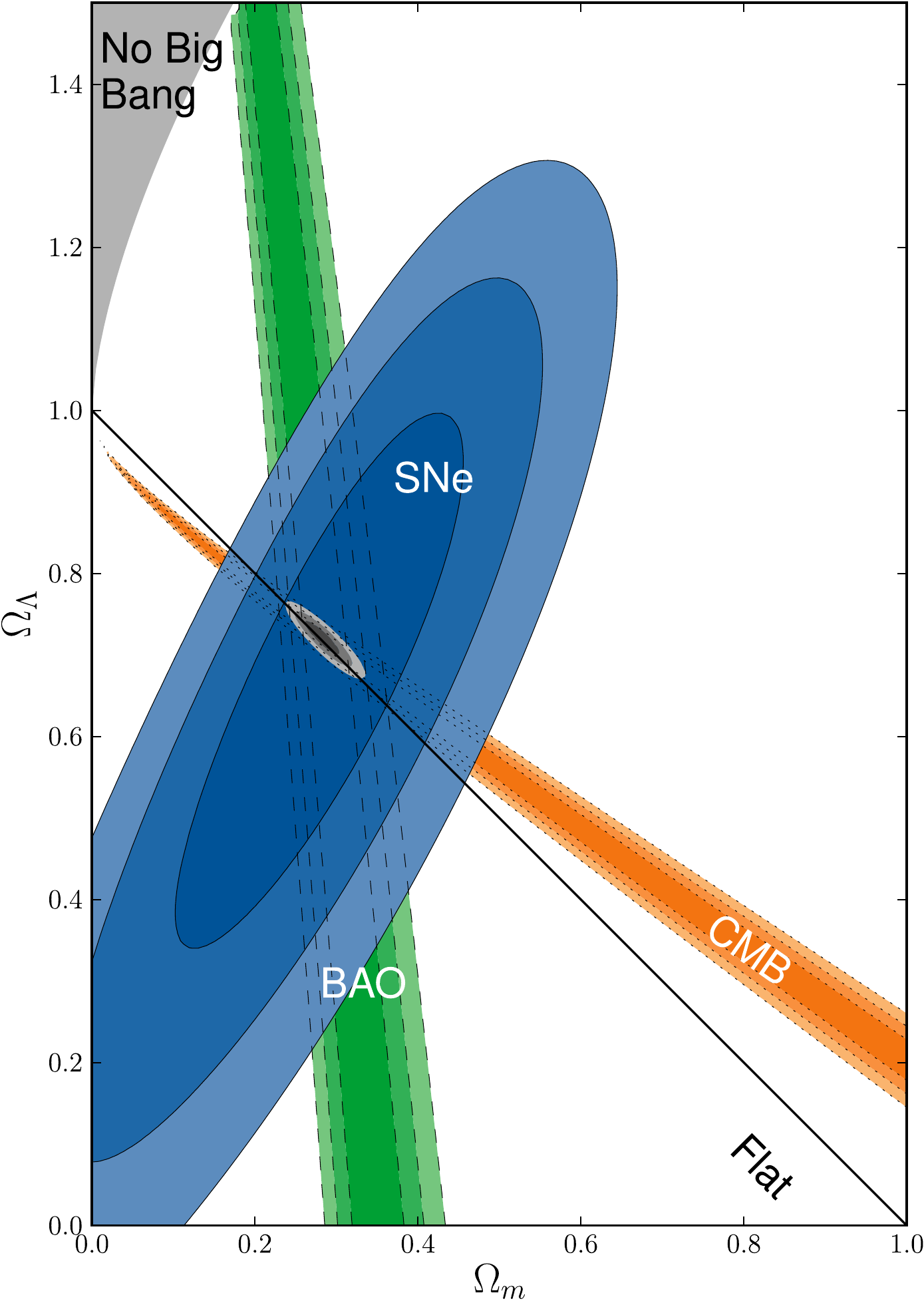}
\raisebox{-0.7cm}{\includegraphics[width=7.5cm, height=7.43cm]{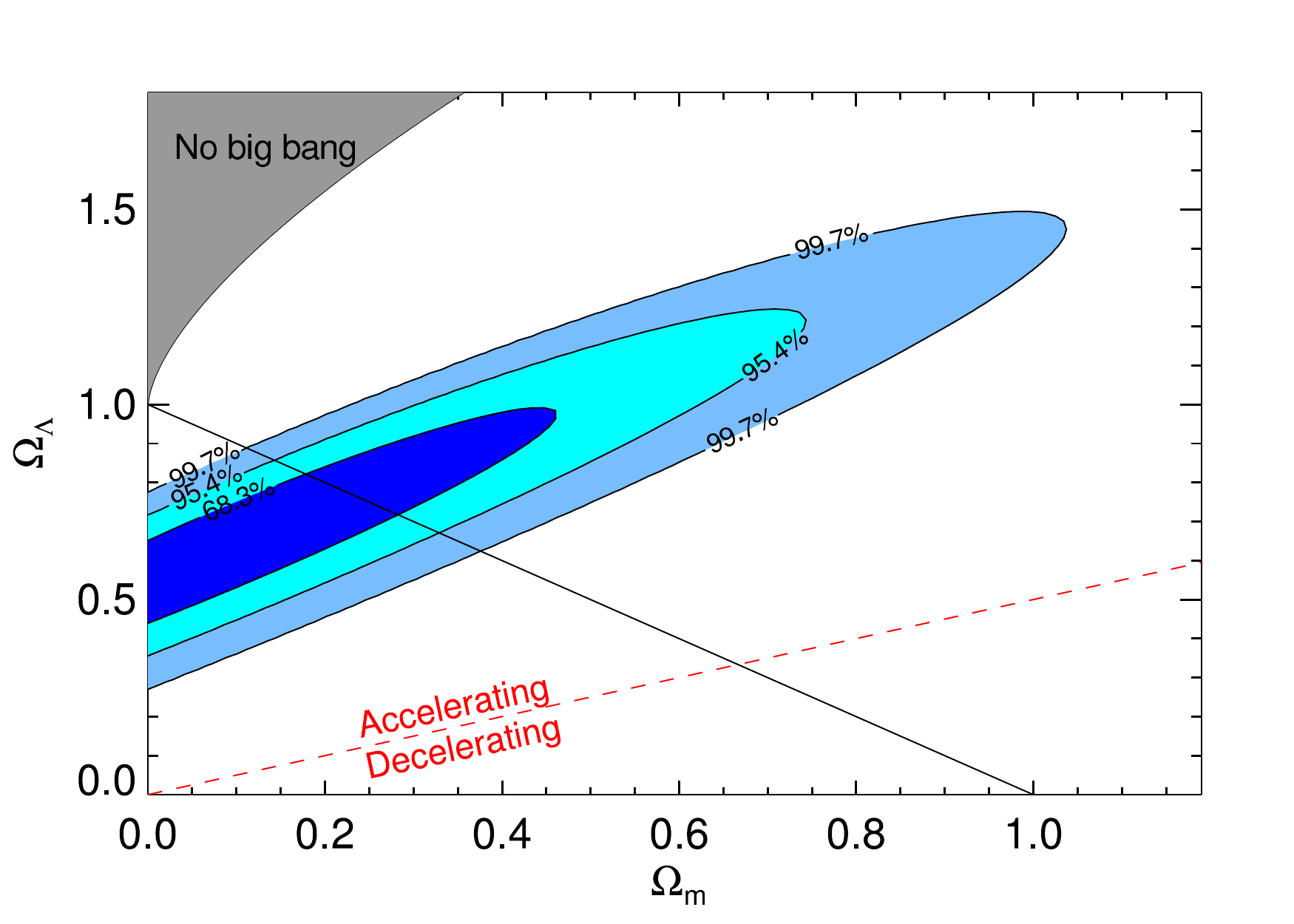}}
\caption[]{\small $68.3\%,\ 94.4\%$ and $99.7\%$ confidence level contour
  in the ($\Omega_m,\Omega_\Lambda$) plane for the SCP sample \cite{suzuki} (left hand
  side) and SNLS 3 years sample \cite{snls} (right hand side). Figures are taken from the references.}
\label{fig2}
\end{center}
\end{figure}

Figures \ref{fig2} show the confidence level contours in the $\Omega_m$,
$\Omega_\Lambda$ plane for these two sets of supernovae. Both samples using
supernovae alone seem to favour a positive value of $\Omega_k$ (i.e $\Omega_t \dpp= \Omega_m + \Omega_\Lambda >1$)
However these two results are statistically compatible with a flat universe because
of the high degeneracy between $\Omega_m$ and $\Omega_\Lambda$ and no conclusion can be drawn. 

Future supernova experiments, like EUCLID \cite{euclid}, WFIRST \cite{wfirst} or LSST
\cite{lsst} won't be able to overcome this degeneracy.
Using our own simulation \cite{kosmoshow} of a WFIRST like survey
(2000 supernovae up to a redshift of 1.7 and intrinsic dispersion of 0.1) 
we find an accuracy on $\Omega_k$ of about 0.05 (statistical error only).
In principle the LSST survey can do better. Using again our simulation 
program with 50000 supernovae per year up to a redshift of 0.8 
with the same intrinsic dispersion gives a statistical accuracy of 0.02 
(0.007 for 10 years running). Those results are only indicative, 
because -- as is well known --  supernova errors are limited by
systematic effects like evolution with redshift, dust or intrinsic variability. 

To overcome this degeneracy we can combine  supernovae with other probes,
like CMB, BAO and weak lensing. Again we can take these combinations directly from published results because the metric is the same 
and perturbation theory can be used as is.  
Table 1 shows the most up-to-date results on  $\Omega_k$. 
All results are in good agreement with each other  and are compatible with a flat universe. 
However a positive curvature is possible at a $2 \sigma$ level \cite{suzuki} if
time evolution of dark energy equation of state is included in the fitting procedure.

Future experiments like PLANCK \cite{planck} combined with WFMOS
\cite{wfmos} or SKA \cite{ska} would achieve an accuracy on $\Omega_k$ 
 between $0.5\cdot10^{-3}$ and $2\cdot 10^{-3}$ with a constant dark energy equation of state
parameter \cite{combine}. This accuracy is degraded by a factor 10 in case of
evolving dark energy equation of state \cite{zhan}.

A positive curvature parameter  measurement $\Omega_k$ above $10^{-2}$ to $10^{-3}$ will reinforce our hypothesis of a maximally symmetric universe on our past light-cones. On the contrary a negative curvature measurement will rull out this hypothesis.

\begin{table} [htbp]
\begin{center}
\label{results}
\begin{tabular}{|c|c|c|c|} \hline
  Experiments        & Errors  &  Level & Reference \\  \hline
WMAP7+SN1a+BAO	&$-0.0133<\Omega_k <0.0084$ &	$95\%$ CL & \cite{wmap7} \\  \hline
Union2+WMAP7+BAO & $\Omega_k = -0.004 \pm 0.007$ & $1 \sigma $ &
\cite{union2} \\  \hline
SNLS+WMAP7+BAO	& $ \Omega_k = -0.002 \pm 0.006$ & $1 \sigma $ &	\cite{snls} \\  \hline
SNe+BAO+CMB+$H_0$ & $\Omega_k = 0.002 \pm 0.005$ & $1 \sigma $ &
\cite{suzuki} \\  \hline \hline
SNe+BAO+CMB+$H_0$ & $\Omega_k = 0.027^{+0.012}_{-0.011}$ & $1 \sigma $ &	\cite{suzuki} \\  \hline
\end{tabular}
\caption[] {Present constraints on spatial curvature parameter. The last line includes time
  evolution of dark energy equation of state in the fitting procedure.}
\end{center}
\end{table}

\section{Epilogue}

\subsection{Mathematical questions}

From the start, we took our light-cones embedded in maximally symmetric spacetimes. This was convenient for the calculation because it by-passed a general theory of isometry groups of degenerate `metrics', which, to the best of our knowledge, is still missing.

As already mentioned, a classification of space-times admitting  foliations with maximally symmetric light-like leaves is needed in order to decide if Roberson-Walker metrics with $\sigma =0$ or $+1$ are indeed incompatible with the relativistic cosmological principle. If we are lucky, this classification might also exhibit a new spacetime metric to be tested in cosmology. (A classification of space-times admitting  foliations with maximally symmetric space-like leaves is also welcome, as it would clarify `Weyl's principle'.)

Our calculation was local and we do not know under which global conditions our Killing vectors do exponentiate. 
Many cosmological models have space-like leaves that are maximally symmetric only locally. However for pseudo-spheres the theory of space forms is more involved  than for planes and spheres \cite{ratc}. 

\subsection{Conclusions}

Any change of paradigm comes with doubts and challenges. We are not sure that the tests involving perturbations and Boltzmann's equation yield the same results for the two different foliations, the one with space-like and the one with light-like leaves. But we do hope that the light-cone coordinates yielding maximally symmetric leaves on the Robertson-Walker metrics with negative curvature can contribute to a simplified description of cosmological observations.

From the strategic point of view, we retain two pleasant features of the relativistic version of the cosmological principle: two expected road blocks did not happen. 
{\it (i)}
The isometry group of the light-cone came out finite dimensional in our, admittedly pedestrian, approach.
{\it (ii)}
To get rid of simultaneity, we started out with another heresy, that of a privileged observer, us. But at the end privilege and heresy vanished.

From the physical point of view, our conclusion amounts to two disappointments: 
{\it (i)}
 Of course we had hoped that our relativistic version of the cosmological principle lead to metrics different from the well studied Robertson-Walker ones. This is not true. But at least there is a constraint, that of negative space-curvature.
{\it (ii)}
 The second disappointment stems from the age of the authors. No statistically significant, dedicated test of the constraint will become available within our life time. Nevertheless it is not quite excluded that the Hubble diagram might show a statistically significant non-monotonicity in our future. This would rule out the relativistic version of the cosmological principle even without assuming Einstein's equation \cite{st}.

\vskip.5 cm \noindent {\bf Acknowledgement:} It is a pleasure to thank Michael Puschnigg,
Christoph Stephan and Charling Tao for their advice. Part of this work was
done at the Tsighua Center for Astrophysics.


\begin{thebibliography}{10}

  \bibitem{fried}
  	F. G. Friedlander, 
{\it A unique continuation theorem for the wave equation in the exterior of a characteristic cone,} S\'eminaire \'Equations aux d\'eriv\'ees partielles (dit ``Goulaouic-Schwartz'') (1982-1983), Exp. No. 2.
\url{http://www.numdam.org/item?id=SEDP_1982-1983____A2_0};\\
  P.~T.~Chrusciel,
  {\it The existence theorem for the general relativistic Cauchy problem on the light-cone,}
  arXiv:1209.1971 [gr-qc] and references therein.
\bibitem{sour}
J.-M. Souriau, {\it M\'ecanique statistique, groupes de Lie et cosmologie}, in {\bf G\'eom\'etrie Symplectique et Physique Math\'ematique}, \'Editions du Centre National de la Recherche Scientifique (1975) Paris.
\bibitem{ms} 
Ch. Marinoni and H. Steigerwald, {\it Emergent flat de Sitter by relaxing FLRW shear free assumption: A counter  example to Weyl's principle}, to appear.
\bibitem{suzuki} N. Suzuki et al., {\it The Hubble Space Telescope Cluster
    Supernova Survey: V. Improving the Dark Energy Constraints Above $z>1$ and
    Building an Early-Type-Hosted Supernova Sample},
   ApJ 746 (2012)  85. 
   \bibitem{snls} M. Sullivan et al.,{\it SNLS3: Constraints on Dark Energy
    Combining the Supernova Legacy Survey Three Year Data with Other
    Probes},
  Astrophys. J. 737 (2011) 102.
  \bibitem{euclid} J. Amiaux et al., {\it Euclid Mission: building of a Reference
    Survey} SPIE Astronomical Telescopes and Instrumentation Proceeding SPIE8442-32 (2012).
\bibitem{wfirst} J. Green et al., {\it Wide-Field Infrared Survey Telescope
    (WFIRST) Final Report}, Instrumentation and Methods for Astrophysics
  (astro-ph, IM) arXiv:1208.4012.
\bibitem{lsst} Paul A. Abell et al., {\it LSST Science Book, Version 2}, 
 Instrumentation and Methods for Astrophysics (astro-ph, IM) arXiv:0.912.0201.
\bibitem{kosmoshow} A. Tilquin
  http://marwww.in2p3.fr/~tilquin/kosmoshowsc.pro
\bibitem{planck} P. A. R. Ade et al., {\it Planck early results. I. The Planck
    mission} A\&A 536 A1 (2011).
    \bibitem{wfmos} 
    Robert Nichol, {\it Measuring Dark Energy with the Wide-Field Multi-Object
Spectrograph (WFMOS)},  arXiv:astro-ph/0611801.
\bibitem{ska} C. Carilli and S. Rawlings, {\it Science with the Square Kilometre
    Array} New Astronomy Reviews, Vol. 48, Elsevier, December 2004.
\bibitem{combine} Mihran Vardanyan et al., {\it How flat can you get? A model
    comparison perspective on the curvature of the Universe},
  Mon. Not. R. Astron. Soc 397 (2009) 431.
\bibitem{zhan} Zhan et al., {\it Distance, Growth Factor, and Dark Energy Constraints from Photometric Baryon Acoustic Oscillation and Weak Lensing Measurements}, Astrophys. J. 690 (2009) 923. 
\bibitem{wmap7} E. Komatsu et al., {\it Seven-Year Wilkinson Microwave
    Anisotropy Probe (WMAP) Observations: Cosmological Interpretation}, Astrophys. J. Suppl. 192 (2011) 18.
\bibitem{union2} R.~Amanullah, et al., {\it Spectra and HST Light Curves of
    Six Type IA Supernovae at 0.511 $<$ z $<$ 1.12 and the Union2 Compilation},
  ApJ April 9, 2010.
  \bibitem{ratc}
  J. G. Ratcliffe, {\bf
Foundations of Hyperbolic Geometry}, (2. edition)
Graduate Texts in Mathematics (2000)
Springer.
  \bibitem{st}
T. Sch\"ucker and A. Tilquin, {\it
From Hubble diagrams to scale factors},
astro-ph/0506457, A\&A 447 (2006) 413.
\end{thebibliography}
\end{document}